\documentclass[preprint,12pt]{elsarticle}

\usepackage{amssymb}
\journal{Current Opinion in Biotechnology}

\begin{document}

\begin{frontmatter}

%% Title, authors and addresses

%% use the tnoteref command within \title for footnotes;
%% use the tnotetext command for theassociated footnote;
%% use the fnref command within \author or \address for footnotes;
%% use the fntext command for theassociated footnote;
%% use the corref command within \author for corresponding author footnotes;
%% use the cortext command for theassociated footnote;
%% use the ead command for the email address,
%% and the form \ead[url] for the home page:
%% \title{Title\tnoteref{label1}}
%% \tnotetext[label1]{}
%% \author{Name\corref{cor1}\fnref{label2}}
%% \ead{email address}
%% \ead[url]{home page}
%% \fntext[label2]{}
%% \cortext[cor1]{}
%% \affiliation{organization={},
%%             addressline={},
%%             city={},
%%             postcode={},
%%             state={},
%%             country={}}
%% \fntext[label3]{}

\title{Defining the boundaries: challenges and advances in identifying cells in microscopy images}

%% use optional labels to link authors explicitly to addresses:
%% \author[label1,label2]{}
%% \affiliation[label1]{organization={},
%%             addressline={},
%%             city={},
%%             postcode={},
%%             state={},
%%             country={}}
%%
%% \affiliation[label2]{organization={},
%%             addressline={},
%%             city={},
%%             postcode={},
%%             state={},
%%             country={}}

\author[inst1]{Nodar Gogoberidze}
\author[inst1]{Beth A. Cimini}

\affiliation[inst1]{organization={Imaging Platform, Broad Institute},%Department and Organization
            %addressline={415 Main St}, 
            city={Cambridge},
            postcode={02142}, 
            state={Ma},
            country={USA}}

\begin{abstract}
%% Text of abstract
Segmentation, or the outlining of objects within images, is a critical step in the measurement and analysis of cells within microscopy images. While improvements continue to be made in tools that rely on classical methods for segmentation, deep learning-based tools increasingly dominate advances in the technology. Specialist models such as Cellpose continue to improve in accuracy and user-friendliness, and segmentation challenges such as the Multi-Modality Cell Segmentation Challenge continue to push innovation in accuracy across widely-varying test data as well as efficiency and usability. Increased attention on documentation, sharing, and evaluation standards are leading to increased user-friendliness and acceleration towards the goal of a truly universal method.
\end{abstract}

%%Graphical abstract
% \begin{graphicalabstract}
% \includegraphics[width=\textwidth]{graphical_abstract.png}
% \end{graphicalabstract}

%%Research highlights
% \begin{highlights}
% \item Adoption of segmentation models driven by user-friendly tools and documentation
% \item Advancements in novel architectures is driven by the need for a truly general model
% \item Greater interoperability between models and tools is needed
% \end{highlights}

%\begin{keyword}
%% keywords here, in the form: keyword \sep keyword
%cell segmentation \sep single cell detection
%% PACS codes here, in the form: \PACS code \sep code
%\PACS 0000 \sep 1111
%% MSC codes here, in the form: \MSC code \sep code
%% or \MSC[2008] code \sep code (2000 is the default)
%\MSC 0000 \sep 1111
%\end{keyword}

\end{frontmatter}

%% \linenumbers

%% main text
\section{Introduction}
\label{sec:intro}
Segmentation plays a pivotal role in microscopy analysis and refers to the automatic delineation of individual objects (often cells or cellular components) within complex scientific images. It is an important step prior to measuring properties of those biological entities. Approaches for cell segmentation have benefitted from advancements in more general segmentation problems in traditional Computer Vision (CV), Machine Learning (ML), and in recent years Deep Learning (DL) \cite{Hollandi2022-wh,Lucas2021-cx}. Accurate segmentation allows the quantification and analysis of cellular features, such as morphology, staining intensity, and spatial relationships, which capture valuable cellular phenotypes. While computational methods now achieve better-than-human accuracy on a number of specific tasks, in general, given the wide range of cell types, imaging modalities, and experimental conditions, the problem remains an ongoing challenge.

As state-of-the-art (SOTA) methodologies for segmentation have progressed, the community has also tried to provide access to these methods to less-computational users in the form of user-friendly software interfaces and intuitive tools that improve reproducibility. Widespread adoption will require methods with few-or-no tunable parameters, models that are efficient in terms of computational runtime and memory requirements, and an ecosystem of tools for their use. The past two years reviewed here have seen a proliferation of new local and cloud-oriented software and workflows, the adoption of several user-oriented models, and the development of next generation model architectures. We herein review progress in approaches utilizing classical computer vision techniques and specialist deep learning networks, as well as progress towards and current needs related to making high quality accessible generalist networks that will reduce the "time to science" for the broader community.

\section{Progress in classical approaches}
\label{sec:progress_classical}
While advancements in segmentation accuracy are largely driven via deep learning approaches, they are not always a suitable solution, as some require large annotated datasets and interpretability (though often unnecessary for segmentation tasks) can be a challenge \cite{Rudin2019-dz,Karim2023-dp}. We therefore begin with advancements in non-deep machine learning and classic image processing. Kartezio \cite{Cortacero2023-en} is a recent exemplar of non-deep ML, using Cartesian Genetic Programming to combine classic Computer Vision algorithms into a fully interpretable image pipeline for segmentation. It performs comparably to deep learning approaches such as Cellpose \cite{Stringer2021-cs}, StarDist \cite{Schmidt2018-ug} and Mask R-CNN \cite{He2020-sg}, and importantly requires relatively few images to train a computationally-efficient and explainable workflow.

Existing CV tools such as Fiji/ImageJ \cite{Schindelin2012-lu}, CellProfiler \cite{Stirling2021-mi}, and Napari \cite{Ahlers2023-xb} are receiving continuous extensions via the growing ecosystem of plugins and integrations \cite{Haase2022-gw,Hollandi2022-wh,Selzer2023-vj,Weisbart2023-bj,Rueden2022-qe}, allowing these tools to adapt to a broader range of tasks. One example, General Image Analysis of Nuclei-based Images (GIANI) \cite{Barry2022-uf} is a Fiji plugin for segmentation of cells in 3D microscopy images. Similarly, LABKIT \cite{Arzt2022-pj} is a Fiji plugin specifically oriented towards efficient segmentation in large, multi-terabyte, images. Other tools, such as Tonga \cite{Ritchie2022-bb} prioritize ease of installation and customization to a specific task to appeal to non-technical and non-expert users.

\section{Lowering time-to-science in deep learning}
\label{sec:lowering_tts}
\begin{figure}
	\centering 
	\includegraphics[width=\textwidth]{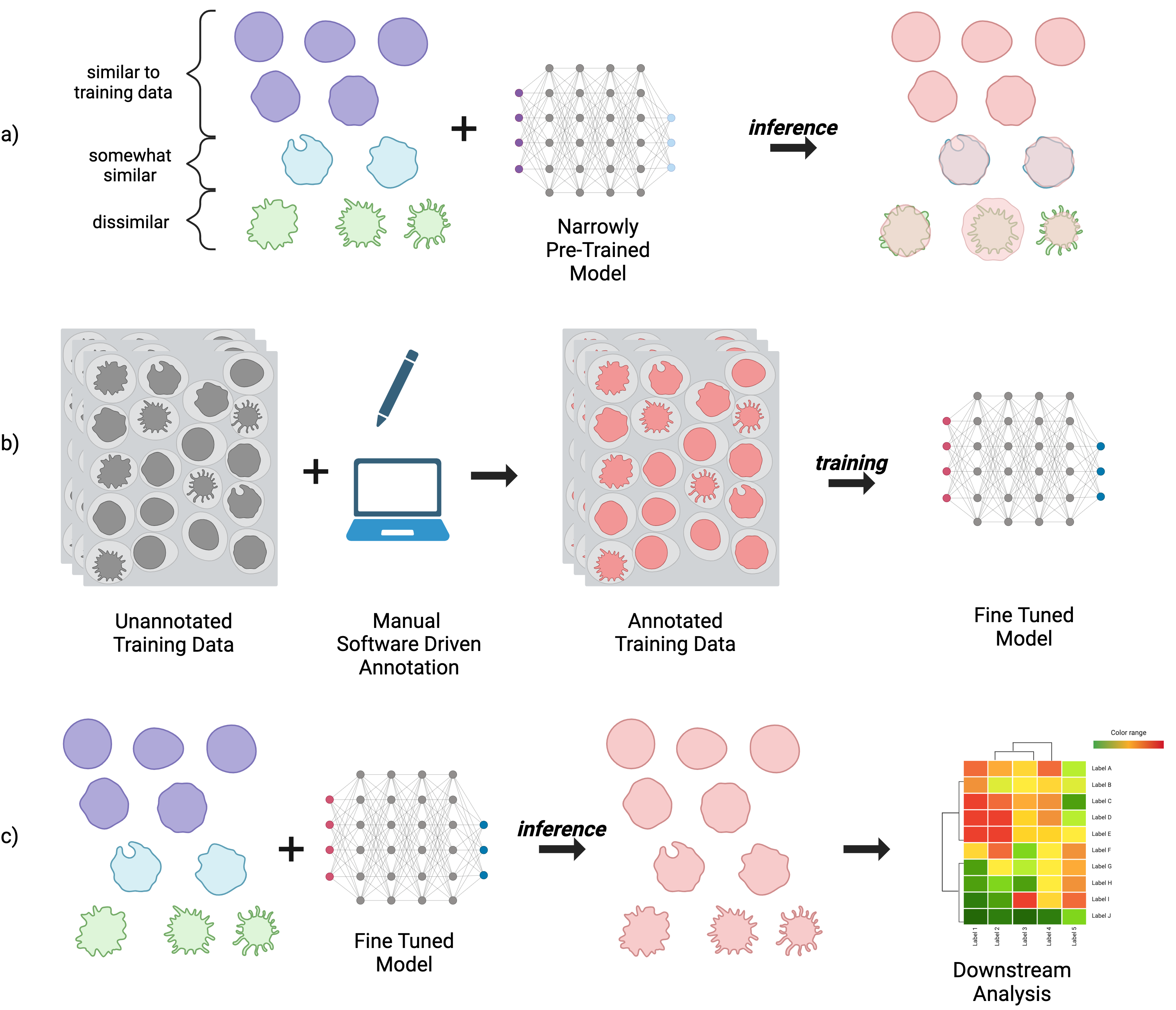}
	\caption{Segmentation and fine tuning process. \textbf{a)} The inference dataset contains a variety of samples; some are similar to the original training data the model was pre-trained on, while others are slightly or very different, leading to low segmentation accuracy. \textbf{b)} New training data matching the characteristics of the full distribution is annotated either manually, through software such as CellProfiler, or a human-in-the-loop model such as Cellpose, and used to fine-tune the model. \textbf{c)} The fine-tuned model produces more accurate segmentations on the inference dataset, which can then be used for downstream tasks.} 
	\label{fig:fig_1}%
\end{figure}

As new deep learning models rapidly emerge, users who want to employ them must surmount computational hurdles. Firstly, installing the models and their corresponding software dependencies can often be technically complex and time-consuming, often involving usage of the Command Line Interface (CLI), compiling source code, navigating environment related conflicts, and other tasks which may confound non-technical users. Secondly, pre-trained models are most effective on the type of data they've been trained on, which may differ from a user's data in a number of ways such as the phenotype of interest, how the biological samples were prepared, the imaging modality (fluorescence, histological stains, phase contrast, etc), and the experimental conditions. The model’s \textit{generality} is how well it’s able to perform across these differences, without loss in segmentation accuracy. If the model’s generality is not sufficient to perform well on the user’s data, it must be fine tuned by feeding in new training data (Figure \ref{fig:fig_1}). The complexity of this task can vary even more dramatically than installation. Depending on the tools made available by the model’s developers, it may involve writing custom code, which can be a time-intensive task even for expert programmers, or may simply require the use of a purpose-built tool.

The less time spent installing, tuning, and configuring models and software, the more bandwidth is available to concentrate on addressing scientific questions. Unfortunately, usability varies wildly across tools and documentation for trained deep learning models, from only the model parameters (weights) without documentation or source code, to models that come with extensive documentation and entire libraries for utilizing the model, including data loading and processing, fine-tuning, and configuration, to models that come with several interfaces including CLI or Graphical User Interfaces (GUI), easy to use installers, and guides or tutorials on how to fine-tune and configure the model. It is no accident that some of the most commonly used networks, discussed below, are those that in addition to high performance have emphasized usability. 

\subsection{Current progress towards useful specialist networks}
\label{sec:progress_specialist}
As shown in Figure \ref{fig:fig_1}, models typically underperform on new, unseen data. Before fine tuning, this data is considered \textit{out of distribution} (OOD). Intuitively, OOD data can be thought of as being drawn from a different distribution than that of the original training set (such as the purple vs green cells in Figure \ref{fig:fig_1}). A shift between the in-distribution training set and OOD data can be referred to as a difference in \textit{style} between the two sets of data; for instance differences in acquisition parameters, staining methods, or imaging modalities \cite{Uhlmann2022-zl,Tian2021-fm}. A process known as \textit{style transfer} can be utilized to address changes in the data distribution by training a model that is able to pixel-wise map an image of one style to that of another, ideally with minimal loss in semantic content \cite{Isola2016-rm}. For instance, a style transfer model can be trained to transform an image of one modality, such as brightfield, to that of another, such as fluorescence; or a model can be trained to transform an image’s annotation mask into the image itself. The nucleAIzer \cite{Hollandi2020-nm} model utilizes the latter approach as a means of achieving greater generalization capabilities, allowing the model to more easily be adapted to OOD data. Although developed over three years ago, it is still one of the top performing models at cell segmentation \cite{Lee2022-zz}. In order to improve usability, a plugin was developed for CellProfiler 3 \cite{McQuin2018-wx} allowing users of the tool to perform inference through the use of a GUI. While the plugin allows ease of use via the CellProfiler interface, manual installation of nucleAIzer and its dependencies are still necessary, which is a challenge for non-computational users. A simple web interface is also available, but requires upload of the data to a central server, limiting use for large batches of images.

StarDist predates nucleAIzer, however iterations of it are still being introduced \cite{Mandal2021-sn,Walter2020-aj}, and general usage remains quite high due to a great deal of time being spent on documenting its usage, as well as making it available across environments and in a variety of graphical tools \cite{Haase2022-gw} such as Fiji/ImageJ, Napari, QuPath \cite{Bankhead2017-zh,Humphries2021-wn}, Icy \cite{De_Chaumont2012-ow}, CellProfiler, and KNIME \cite{Fillbrunn2017-ef}. While a variety of pretrained variants are available, it is still limited to specific modalities (such as fluorescence or histology stains) and even with fine-tuning is oriented to segmenting objects which are \textit{star-convex}---shapes where line segments can be draw from any point along the border to some single interior point---which makes it a poor choice for very irregular cell shapes or neurons, since shapes with very large bends or curves may not be star-convex.

A primary goal of Cellpose was to develop a generalist model by training on a large dataset of manually segmented images from a variety of modalities. Its preprocessing method, focusing on transforming input data to spatial gradients, allows it to generalize to a larger variety of shapes. While the architecture was developed for generalist purposes, fine-tuning is still often necessary. Cellpose 2.0 \cite{Pachitariu2022-ve} was introduced as a package that included several pre-trained models, a human-in-the-loop pipeline for fine-tuning custom models with small datasets, and an improved set of graphical software tools to aid in its usage. Omnipose \cite{Cutler2022-mx} extends Cellpose to work better on elongated cells common in bacteria by adding distance field prediction similar to StarDist. It is similarly well documented, packaged, and provides a number of interfaces in the form of a library, command line interface, and GUI. While it allows for training, it is not available in Cellpose 2.0, and thus is not focused on human-in-the-loop finetuning. Cellpose and Cellpose 2.0 include a custom GUI, and extensive documentation; broad usage is further supported by their availability via plugins from many of the same tools as StarDist.

Mesmer is a deep learning pipeline trained on the largest public tissue data set of annotated nuclei and whole cells, TissueNet \cite{Greenwald2022-gm}. It provides access to a remotely hosted instance model through several interfaces, including a web portal and plugins for Fiji/ImageJ and QuPath. It also provides a Docker container (described below, in \ref{sec:models_fair}) to run the model in a self hosted manner, with access provided through a Jupyter Notebook or CLI.

\section{A Vision for the State-Of-The-Art}
\label{sec:vision_sota}
Many of those working on deep learning for image segmentation are aiming to create a truly generalist model, often referred to as a foundation model, capable of greater-than-human accuracy, across the wide variety of imaging modalities, in a parameter-free manner \cite{Ma2023-bh,Laine2021-qu,Ma2023-zo}. For biologists, this would mean a model that has no need for manual annotation or fine tuning, preferably in a form factor that is easily accessible, configurable, and invocable. Although there is still much progress to be made on this front, a number of architectural developments can be highlighted as mile markers along the path.

\subsection{Dataset Availability}
\label{sec:dset_availability}
The effectiveness of Deep Learning models relies on sufficient similarity between the training data and the user’s own data. The specific criteria for what constitutes “sufficient similarity” will naturally differ based on the methods and architectural choices employed by the model, but it therefore follows that creating foundation models (capable of segmenting a wide variety of biological images) will require a diverse and comprehensive corpus of training data to ensure that the segmentation model can generalize effectively. The dataset should encompass a broad spectrum of microscope modalities, across a variety of imaging conditions, and include many distinct cell types.

The volume of publicly available datasets is ever increasing, particularly driven from the development of specialist models and challenges wherein teams compete on segmentation oriented tasks. TissueNet is the largest collection of annotated tissue images, while LIVECell \cite{Edlund2021-lw} is the largest collection of high-quality, manually annotated, and expert-validated phase-contrast images. The Cell Tracking Challenge (CTC) \cite{Maska2023-db} is an ongoing benchmark and reference in cell segmentation and tracking algorithms, which in recent years has extended the available benchmarks with the Cell Segmentation Benchmark (CSB). The Multi-Modality Cell Segmentation Challenge (MMCSC) \cite{Ma2023-zo} consolidated a modestly sized labeled dataset with a particular emphasis on diversity in modalities.

Challenges stand as an excellent pointer for future progress in a given area; trends among the top-ranking team’s architectures and techniques often form the basis of future implementations available to the wider field. For instance, data augmentation—where existing training data is perturbed and transformed in various ways, such as rotations, scaling, intensity adjustments, or the infusion of random noise—was highlighted in the MMCSC as a particularly important feature in pre-training top performing models, aiding them in their generalizability. Entirely synthetic datasets are also often useful, as demonstrated by their inclusion in a subset of the CTC datasets, and in the development of frameworks for their generation \cite{Dey2023-ei}.

\subsection{Next Generation Models}
\label{sec:next_gen_models}
Many architectures are being researched and explored in the quest for ever more general and robust models, as demonstrated in the MMCSC. The KIT-GE model \cite{Scherr2020-ml} is among the top-3 performing models of CTC CSB, and was therefore used as one of the baselines in the MMCSC alongside Cellpose, Cellpose 2.0, and OmniPose \cite{Pachitariu2022-ve,Cutler2022-mx,Stringer2021-cs}. Analysis of the top-ranking solutions in the MMCSC shows that choices in backbone networks are particularly important for next generation models. While U-Net inspired architectures form the basis of the widely used contemporary models such as StarDist, Cellpose, and KIT-GE, the winning solutions in MMCSC employ backbones such as SegFormer \cite{Xie2021-tr}, ConvNeXt \cite{Liu2022-ns}, and ResNeXt \cite{Xie2016-fp}. CTC reports that segmentation performance increases with techniques such as self-configured neural networks (e.g. nnU-Net) \cite{Isensee2021-wd}, neural architecture search, and multi-branch prediction.

The generalist capabilities provided by Cellpose 2.0 rely on fine-tuning, which may cause the model to suffer from severe loss of performance on tasks outside of the fine-tuning dataset \cite{Ma2023-zo}. The top performing model \cite{Lee2022-by} of the MMCSC was able to outperform the pre-trained generalist Cellpose and Omnipose models, as well as a Cellpose 2.0 model fine-tuned on the challenge’s training data. The testing set included images which were distinct from the training data, and sourced from new biological experiments, meaning successful models needed to show a strong ability to generalize across data without additional fine-tuning.

While the winning solution of MMCSC and its associated code is available on GitHub, it remains to be seen whether it or any of the top-ranking models will be available with documentation and interface tools in a way comparable to StarDist, Cellpose 2.0, or Omnipose. If so, we may be one step closer to a truly general, easy to set up, easy to use, one-click segmentation model, with no additional tuning. Short of that, alternative interfaces for model configuration may come to prominence in the form of dialog-driven LLMs \cite{Royer2023-ft,Wu2023-au}. It also as yet unclear whether future enhancements will be driven primarily by transformer architectures \cite{Li2023-jw,Ma2023-bh}, whether advancements in convolutional networks \cite{Liu2022-ns} will keep pace, as demonstrated by the second and third place solutions in the MMCSC, or whether hybrid approaches will dominate \cite{Gao2021-ft,Wang2022-jp}.

It is also of great research and commercial interest to develop foundation models. Meta AI Research recently released a family of foundation models for segmentation, referred to as the Segment Anything Model \cite{Kirillov2023-qn} (SAM), the largest of which was trained on 1.1 billion high-quality segmentation labels, across 11 million high-resolution images. While the images included in the dataset were mostly photographs of natural scenes, it did include a small number of microscopy images taken from the 2018 Data Science Bowl \cite{Caicedo2019-dc}. In short order, Segment Anything for Microscopy \cite{Archit2023-kl} was developed, in which SAM was extended to generalize across many imaging modalities by fine-tuning the original model using a variety of datasets. Important to SAM’s architecture is its interactive segmentation capabilities, where a subset of the user’s data is first annotated with a small amount of either point annotations or rectangular bounding boxes. Annotations of this type are significantly less time-consuming than pixel-level mask annotations, and provide SAM with enough guidance to output full segmentation masks on the user’s dataset. Segment Anything for Microscopy therefore adopts this capability and includes a Napari plugin for interactive and automatic segmentation. There is a mechanism for automatic segmentation, however in order to get generalist accuracy above that of Cellpose, some manual annotations must be made.

\subsection{Making models FAIR}
\label{sec:models_fair}
Alongside progress in model development, there has been a greater push towards the dissemination of models such that they are Findable, Accessible, Interoperable, and Reusable (FAIR) \cite{Paul-Gilloteaux2023-cw,Kemmer2023-ko,Wilkinson2016-sq}. Not only should models be available on publicly accessible platforms, but the associated code for asset loading, data pre-processing, data post-processing, model training and model inference, should also be made available in a well packaged and documented form. Container platforms such as Docker \cite{Merkel2014-xs} can alleviate many installation and setup complexities, providing an isolated and controlled environment in which software is installed, and a pre-configured installation process for the target software. This “containerization” of software and its dependencies dramatically decreases the barriers for reuse. In addition, making models readily accessible, configurable, and tunable in a low or no code manner via interactive code notebooks or graphical user interfaces encourages broader adoption of SOTA models.

The CTC proposed guidelines for algorithm developers to make their workflows both available and reproducible; while currently optional, they will be mandatory in the future. At minimum the source code should be available on a public repository, and contain clear instructions for installing dependencies, initializing the model, loading weights, and training with new data. They also pushed for source code to be available via notebooks such as Jupyter and Google Colab. In the same vein, the MMCSC required all participants to place their solutions in Docker containers; the winning teams have made these available on public image registries and also made their algorithms publicly available on GitHub alongside processing source code. In addition the top three solutions were encouraged to develop Napari \cite{Ahlers2023-xb} plugins.

A missing component in full adherence to FAIR principles is the interoperability of models. While research and development in model architectures is healthy and vital, there is no agreed upon specification on the inputs and outputs of models. The difference in the model outputs between e.g. StarDist and Cellpose is stark, and the post-processing that is needed is correspondingly distinct. While the outputs are necessary byproducts of the model architectures, the lack of a standard makes interoperability with existing tools difficult as custom code needs to be written to mirror the post-processing steps.

\subsection{Making models efficient}
\label{sec:models_efficient}
Models vary widely in terms of algorithmic efficiency, which will affect their adoption, especially in low-resource settings. While some challenges such as the CTC emphasize segmentation accuracy alone, others such as MMCSC evaluate efficiency as an explicit criterion, and the top performing models had good tradeoffs between accuracy and efficiency in runtime and memory usage. There are additional efforts in bringing model optimization tools to the bioimage community \cite{Zhou2023-dz,Saraiva2023-nf}, as well as reducing runtime of bioimage analysis pipelines in general \cite{Saraiva2023-nf,Haase2020-pd}. Efficient models, and model optimization tools will become increasingly important for training and inference tasks in local desktop and web-based tools \cite{Lucas2021-cx,Ouyang2023-sf}, especially in contexts where moving data to the cloud is not viable or allowed.

\section{Improving tool access and availability}
\label{sec:improve_tool_access}
Though local-first software and algorithms remain important to grow and maintain, cloud oriented tools and resources for bioimage analysis are becoming more prevalent and easier to use as demand for large-scale cloud-based workflows increases. In addition to providing large storage capacity and high performance hardware, cloud-based tools increase the availability and accessibility of models, and often move the technical complexity away from the end user.

Notebooks allow code, explanatory text and interactive elements to live together in a single package, providing alternatives or complements to libraries, documentation and GUI interfaces. ZeroCostDL4Mic \cite{Von_Chamier2021-fp} provides rigorously documented and annotated code notebooks with pre-written code which can be customized for specific workflows through the exposed settings. A major benefit to these notebooks is that they can be deployed either locally or on the Google Colab platform, which eases hardware requirements and allows running moderately sized workloads for free. Behind the scenes, a container is initialized in the cloud and the installation occurs via pre-configured installation scripts contained within the notebooks.

Beyond notebooks, other tools provide a larger degree of customization and control, albeit at the expense of additional complexity. The BioContainers project is an open-source and community-driven framework which provides cloud resources for defining, building, and distributing containers for biological tools \cite{Da_Veiga_Leprevost2017-cc}. The BioContainers Registry was developed with FAIR principles in mind, and provides both web and RESTful API interfaces to search for bioinformatics tools \cite{Bai2021-vb}. BIAFLOWS \cite{Rubens2020-fr} is a community-driven, open-source web platform that allows deployment of and access to a wide variety of reproducible image analysis workflows. The platform provides a framework to import data, encapsulate workflows in container images, batch process data, visualize data, and assess performance using widely accepted benchmark metrics on a large collection of public datasets. BioImageIT \cite{Prigent2022-xg} is a more recent, plugin oriented, workflow tool for data management and analysis. It has a unique emphasis on reconciling existing data management and data analysis tools, and although run locally has the ability to tap into remote data stores and job runners.

The BioImage Model Zoo \cite{Ouyang2022-zw} provides a community-driven repository for pre-trained deep learning models and promotes a standard model description format for describing metadata. Community partners can work with the BioImage Model Zoo to support execution of the models and include many common bioimage tools. In addition, model execution can be performed via the BioEngine application framework, on top of the ImJoy plugin framework \cite{Ouyang2019-yw}, allowing inference both on the BioImage Model Zoo web application, and other web applications using the client ImJoy software. Behind the scenes, multiple containers are being run and managed with a container orchestration tool called Kubernetes.

For moderately more technical users, who are comfortable with using tools for deploying their own container orchestration workloads, there are some additional options. DeepCell Kiosk \cite{Bannon2021-oz} is a cloud-native tool for dynamic scaling of image analysis workflows, utilizing Kubernetes orchestration similarly to the BioEngine inference engine. The tool is managed from several interfaces including a web portal and Fiji plugin. Distributed-Something \cite{Weisbart2023-mq} takes a script-based approach to scale and distribute arbitrary containerized jobs on AWS, automatically configuring the AWS infrastructure for container orchestration, monitoring, and data handling. It runs the work in a cost effective manner, and cleans up the infrastructure when the work has been completed.

\section{Conclusion}
\label{sec:conclusion}
The landscape of segmentation algorithms, enabling tools, workflow management systems, repositories, benchmarks, and challenges is constantly shifting. This very active landscape makes it all the more important to create community standards for reporting on methods and robust  segmentation quality metrics, on which there has been recent guidance \cite{Schmied2023-xy,Hirling2023-eg,Maier-Hein2022-eb,Laine2021-qu}. While there is still much work to do, the past two years have seen essential strides made in democratizing the use of advanced segmentation methods through user-friendly interfaces and improved documentation. Integrating tools and scaling up reproducible workflows fosters a more collaborative and robust ecosystem; these continuing trends will empower researchers from diverse backgrounds to collectively explore the intricate universe of single-cell biology, ultimately accelerating the pace of discovery and innovation in this vital field of study.

 \bibliographystyle{elsarticle-num} 
 \bibliography{cas-refs}

\begin{thebibliography}{10}
\expandafter\ifx\csname url\endcsname\relax
  \def\url#1{\texttt{#1}}\fi
\expandafter\ifx\csname urlprefix\endcsname\relax\def\urlprefix{URL }\fi
\expandafter\ifx\csname href\endcsname\relax
  \def\href#1#2{#2} \def\path#1{#1}\fi

\bibitem{Hollandi2022-wh}
R.~Hollandi, N.~Moshkov, L.~Paavolainen, E.~Tasnadi, F.~Piccinini, P.~Horvath, Nucleus segmentation: towards automated solutions, Trends Cell Biol. 32~(4) (2022) 295--310.

\bibitem{Lucas2021-cx}
A.~M. Lucas, P.~V. Ryder, B.~Li, B.~A. Cimini, K.~W. Eliceiri, A.~E. Carpenter, Open-source deep-learning software for bioimage segmentation, Mol. Biol. Cell 32~(9) (2021) 823--829.

\bibitem{Rudin2019-dz}
C.~Rudin, Stop explaining black box machine learning models for high stakes decisions and use interpretable models instead, Nat Mach Intell 1~(5) (2019) 206--215.

\bibitem{Karim2023-dp}
M.~R. Karim, T.~Islam, {Shajalal}, O.~Beyan, C.~Lange, M.~Cochez, D.~Rebholz-Schuhmann, S.~Decker, Explainable {AI} for bioinformatics: Methods, tools and applications, Brief. Bioinform. 24~(5) (2023) bbad236.

\bibitem{Cortacero2023-en}
K.~Cortacero, B.~McKenzie, S.~M{\"u}ller, R.~Khazen, F.~Lafouresse, G.~Corsaut, N.~Van~Acker, F.-X. Frenois, L.~Lamant, N.~Meyer, B.~Vergier, D.~G. Wilson, H.~Luga, O.~Staufer, M.~L. Dustin, S.~Valitutti, S.~Cussat-Blanc, Kartezio: Evolutionary design of explainable pipelines for biomedical image analysis (Feb. 2023).
\newblock \href {http://arxiv.org/abs/2302.14762} {\path{arXiv:2302.14762}}.

\bibitem{Stringer2021-cs}
C.~Stringer, T.~Wang, M.~Michaelos, M.~Pachitariu, Cellpose: a generalist algorithm for cellular segmentation, Nat. Methods 18~(1) (2021) 100--106.

\bibitem{Schmidt2018-ug}
U.~Schmidt, M.~Weigert, C.~Broaddus, G.~Myers, Cell detection with {Star-Convex} polygons, in: Medical Image Computing and Computer Assisted Intervention -- {MICCAI} 2018, Springer International Publishing, 2018, pp. 265--273.

\bibitem{He2020-sg}
K.~He, G.~Gkioxari, P.~Dollar, R.~Girshick, Mask {R-CNN}, IEEE Trans. Pattern Anal. Mach. Intell. 42~(2) (2020) 386--397.

\bibitem{Schindelin2012-lu}
J.~Schindelin, I.~Arganda-Carreras, E.~Frise, V.~Kaynig, M.~Longair, T.~Pietzsch, S.~Preibisch, C.~Rueden, S.~Saalfeld, B.~Schmid, J.-Y. Tinevez, D.~J. White, V.~Hartenstein, K.~Eliceiri, P.~Tomancak, A.~Cardona, Fiji: an open-source platform for biological-image analysis, Nat. Methods 9~(7) (2012) 676--682.

\bibitem{Stirling2021-mi}
D.~R. Stirling, M.~J. Swain-Bowden, A.~M. Lucas, A.~E. Carpenter, B.~A. Cimini, A.~Goodman, {CellProfiler} 4: improvements in speed, utility and usability, BMC Bioinformatics 22~(1) (2021) 433.

\bibitem{Ahlers2023-xb}
J.~Ahlers, D.~Althviz~Mor{\'e}, O.~Amsalem, A.~Anderson, G.~Bokota, P.~Boone, J.~Bragantini, G.~Buckley, A.~Burt, M.~Bussonnier, A.~Can~Solak, C.~Caporal, D.~Doncila~Pop, K.~Evans, J.~Freeman, L.~Gaifas, C.~Gohlke, K.~Gunalan, H.~Har-Gil, M.~Harfouche, K.~I.~S. Harrington, V.~Hilsenstein, K.~Hutchings, T.~Lambert, J.~Lauer, G.~Lichtner, Z.~Liu, L.~Liu, A.~Lowe, L.~Marconato, S.~Martin, A.~McGovern, L.~Migas, N.~Miller, H.~Mu{\~n}oz, J.-H. M{\"u}ller, C.~Nauroth-Kre{\ss}, J.~Nunez-Iglesias, C.~Pape, K.~Pevey, G.~Pe{\~n}a-Castellanos, A.~Pierr{\'e}, J.~Rodr{\'\i}guez-Guerra, D.~Ross, L.~Royer, C.~T. Russell, G.~Selzer, P.~Smith, P.~Sobolewski, K.~Sofiiuk, N.~Sofroniew, D.~Stansby, A.~Sweet, W.-M. Vierdag, P.~Wadhwa, M.~Weber~Mendon{\c c}a, J.~Windhager, P.~Winston, K.~Yamauchi, napari: a multi-dimensional image viewer for python (2023).

\bibitem{Haase2022-gw}
R.~Haase, E.~Fazeli, D.~Legland, M.~Doube, S.~Culley, I.~Belevich, E.~Jokitalo, M.~Schorb, A.~Klemm, C.~Tischer, A hitchhiker's guide through the bio-image analysis software universe, FEBS Lett. 596~(19) (2022) 2472--2485.

\bibitem{Selzer2023-vj}
G.~J. Selzer, C.~T. Rueden, M.~C. Hiner, E.~L. Evans, 3rd, K.~I.~S. Harrington, K.~W. Eliceiri, napari-imagej: {ImageJ} ecosystem access from napari, Nat. Methods (Aug. 2023).

\bibitem{Weisbart2023-bj}
E.~Weisbart, C.~Tromans-Coia, B.~Diaz-Rohrer, D.~R. Stirling, F.~Garcia-Fossa, R.~A. Senft, M.~C. Hiner, M.~B. de~Jesus, K.~W. Eliceiri, B.~A. Cimini, {CellProfiler} plugins - an easy image analysis platform integration for containers and python tools, J. Microsc. (Sep. 2023).

\bibitem{Rueden2022-qe}
C.~T. Rueden, M.~C. Hiner, E.~L. Evans, 3rd, M.~A. Pinkert, A.~M. Lucas, A.~E. Carpenter, B.~A. Cimini, K.~W. Eliceiri, {PyImageJ}: A library for integrating {ImageJ} and python, Nat. Methods 19~(11) (2022) 1326--1327.

\bibitem{Barry2022-uf}
D.~J. Barry, C.~Gerri, D.~M. Bell, R.~D'Antuono, K.~K. Niakan, {GIANI--open-source} software for automated analysis of {3D} microscopy images, J. Cell Sci. 135~(10) (2022) jcs259511.

\bibitem{Arzt2022-pj}
M.~Arzt, J.~Deschamps, C.~Schmied, T.~Pietzsch, D.~Schmidt, P.~Tomancak, R.~Haase, F.~Jug, {LABKIT}: Labeling and segmentation toolkit for big image data, Frontiers in Computer Science 4 (2022).

\bibitem{Ritchie2022-bb}
A.~Ritchie, S.~Laitinen, P.~Katajisto, J.~I. Englund, ``tonga'': A novel toolbox for straightforward bioimage analysis, Frontiers in Computer Science 4 (2022).

\bibitem{Uhlmann2022-zl}
V.~Uhlmann, L.~Donati, D.~Sage, A practical guide to supervised deep learning for bioimage analysis: Challenges and good practices, IEEE Signal Process. Mag. 39~(2) (2022) 73--86.

\bibitem{Tian2021-fm}
J.~Tian, Y.-C. Hsu, Y.~Shen, H.~Jin, Z.~Kira, Exploring covariate and concept shift for detection and calibration of {Out-of-Distribution} data (Oct. 2021).
\newblock \href {http://arxiv.org/abs/2110.15231} {\path{arXiv:2110.15231}}.

\bibitem{Isola2016-rm}
P.~Isola, J.-Y. Zhu, T.~Zhou, A.~A. Efros, Image-to-image translation with conditional adversarial networks (2016) 1125--1134\href {http://arxiv.org/abs/1611.07004} {\path{arXiv:1611.07004}}.

\bibitem{Hollandi2020-nm}
R.~Hollandi, A.~Szkalisity, T.~Toth, E.~Tasnadi, C.~Molnar, B.~Mathe, I.~Grexa, J.~Molnar, A.~Balind, M.~Gorbe, M.~Kovacs, E.~Migh, A.~Goodman, T.~Balassa, K.~Koos, W.~Wang, J.~C. Caicedo, N.~Bara, F.~Kovacs, L.~Paavolainen, T.~Danka, A.~Kriston, A.~E. Carpenter, K.~Smith, P.~Horvath, {nucleAIzer}: A parameter-free deep learning framework for nucleus segmentation using image style transfer, Cell Syst 10~(5) (2020) 453--458.e6.

\bibitem{Lee2022-zz}
M.~Y. Lee, J.~S. Bedia, S.~S. Bhate, G.~L. Barlow, D.~Phillips, W.~J. Fantl, G.~P. Nolan, C.~M. Sch{\"u}rch, {CellSeg}: a robust, pre-trained nucleus segmentation and pixel quantification software for highly multiplexed fluorescence images, BMC Bioinformatics 23~(1) (2022) 46.

\bibitem{McQuin2018-wx}
C.~McQuin, A.~Goodman, V.~Chernyshev, L.~Kamentsky, B.~A. Cimini, K.~W. Karhohs, M.~Doan, L.~Ding, S.~M. Rafelski, D.~Thirstrup, W.~Wiegraebe, S.~Singh, T.~Becker, J.~C. Caicedo, A.~E. Carpenter, {CellProfiler} 3.0: Next-generation image processing for biology, PLoS Biol. 16~(7) (2018) e2005970.

\bibitem{Mandal2021-sn}
S.~Mandal, V.~Uhlmann, {SplineDist}: Automated cell segmentation with spline curves (Jan. 2021).

\bibitem{Walter2020-aj}
F.~C. Walter, S.~Damrich, F.~A. Hamprecht, {MultiStar}: Instance segmentation of overlapping objects with {Star-Convex} polygons (Nov. 2020).
\newblock \href {http://arxiv.org/abs/2011.13228} {\path{arXiv:2011.13228}}.

\bibitem{Bankhead2017-zh}
P.~Bankhead, M.~B. Loughrey, J.~A. Fern{\'a}ndez, Y.~Dombrowski, D.~G. McArt, P.~D. Dunne, S.~McQuaid, R.~T. Gray, L.~J. Murray, H.~G. Coleman, J.~A. James, M.~Salto-Tellez, P.~W. Hamilton, {QuPath}: Open source software for digital pathology image analysis, Sci. Rep. 7~(1) (2017) 16878.

\bibitem{Humphries2021-wn}
M.~P. Humphries, P.~Maxwell, M.~Salto-Tellez, {QuPath}: The global impact of an open source digital pathology system, Comput. Struct. Biotechnol. J. 19 (2021) 852--859.

\bibitem{De_Chaumont2012-ow}
F.~de~Chaumont, S.~Dallongeville, N.~Chenouard, N.~Herv{\'e}, S.~Pop, T.~Provoost, V.~Meas-Yedid, P.~Pankajakshan, T.~Lecomte, Y.~Le~Montagner, T.~Lagache, A.~Dufour, J.-C. Olivo-Marin, Icy: an open bioimage informatics platform for extended reproducible research, Nat. Methods 9~(7) (2012) 690--696.

\bibitem{Fillbrunn2017-ef}
A.~Fillbrunn, C.~Dietz, J.~Pfeuffer, R.~Rahn, G.~A. Landrum, M.~R. Berthold, {KNIME} for reproducible cross-domain analysis of life science data, J. Biotechnol. 261 (2017) 149--156.

\bibitem{Pachitariu2022-ve}
M.~Pachitariu, C.~Stringer, Cellpose 2.0: how to train your own model, Nat. Methods 19~(12) (2022) 1634--1641.

\bibitem{Cutler2022-mx}
K.~J. Cutler, C.~Stringer, T.~W. Lo, L.~Rappez, N.~Stroustrup, S.~Brook~Peterson, P.~A. Wiggins, J.~D. Mougous, Omnipose: a high-precision morphology-independent solution for bacterial cell segmentation, Nat. Methods 19~(11) (2022) 1438--1448.

\bibitem{Greenwald2022-gm}
N.~F. Greenwald, G.~Miller, E.~Moen, A.~Kong, A.~Kagel, T.~Dougherty, C.~C. Fullaway, B.~J. McIntosh, K.~X. Leow, M.~S. Schwartz, C.~Pavelchek, S.~Cui, I.~Camplisson, O.~Bar-Tal, J.~Singh, M.~Fong, G.~Chaudhry, Z.~Abraham, J.~Moseley, S.~Warshawsky, E.~Soon, S.~Greenbaum, T.~Risom, T.~Hollmann, S.~C. Bendall, L.~Keren, W.~Graf, M.~Angelo, D.~Van~Valen, Whole-cell segmentation of tissue images with human-level performance using large-scale data annotation and deep learning, Nat. Biotechnol. 40~(4) (2022) 555--565.

\bibitem{Ma2023-bh}
J.~Ma, B.~Wang, Towards foundation models of biological image segmentation, Nat. Methods 20~(7) (2023) 953--955.

\bibitem{Laine2021-qu}
R.~F. Laine, I.~Arganda-Carreras, R.~Henriques, G.~Jacquemet, Avoiding a replication crisis in deep-learning-based bioimage analysis, Nat. Methods 18~(10) (2021) 1136--1144.

\bibitem{Ma2023-zo}
J.~Ma, R.~Xie, S.~Ayyadhury, C.~Ge, A.~Gupta, R.~Gupta, S.~Gu, Y.~Zhang, G.~Lee, J.~Kim, W.~Lou, H.~Li, E.~Upschulte, T.~Dickscheid, J.~G. de~Almeida, Y.~Wang, L.~Han, X.~Yang, M.~Labagnara, S.~J. Rahi, C.~Kempster, A.~Pollitt, L.~Espinosa, T.~Mignot, J.~M. Middeke, J.-N. Eckardt, W.~Li, Z.~Li, X.~Cai, B.~Bai, N.~F. Greenwald, D.~Van~Valen, E.~Weisbart, B.~A. Cimini, Z.~Li, C.~Zuo, O.~Br{\"u}ck, G.~D. Bader, B.~Wang, The multi-modality cell segmentation challenge: Towards universal solutions (Aug. 2023).
\newblock \href {http://arxiv.org/abs/2308.05864} {\path{arXiv:2308.05864}}.

\bibitem{Edlund2021-lw}
C.~Edlund, T.~R. Jackson, N.~Khalid, N.~Bevan, T.~Dale, A.~Dengel, S.~Ahmed, J.~Trygg, R.~Sj{\"o}gren, {LIVECell---A} large-scale dataset for label-free live cell segmentation, Nat. Methods 18~(9) (2021) 1038--1045.

\bibitem{Maska2023-db}
M.~Ma{\v s}ka, V.~Ulman, P.~Delgado-Rodriguez, E.~G{\'o}mez-de Mariscal, T.~Ne{\v c}asov{\'a}, F.~A. Guerrero~Pe{\~n}a, T.~I. Ren, E.~M. Meyerowitz, T.~Scherr, K.~L{\"o}ffler, R.~Mikut, T.~Guo, Y.~Wang, J.~P. Allebach, R.~Bao, N.~M. Al-Shakarji, G.~Rahmon, I.~E. Toubal, K.~Palaniappan, F.~Lux, P.~Matula, K.~Sugawara, K.~E.~G. Magnusson, L.~Aho, A.~R. Cohen, A.~Arbelle, T.~Ben-Haim, T.~R. Raviv, F.~Isensee, P.~F. J{\"a}ger, K.~H. Maier-Hein, Y.~Zhu, C.~Ederra, A.~Urbiola, E.~Meijering, A.~Cunha, A.~Mu{\~n}oz-Barrutia, M.~Kozubek, C.~Ortiz-de Sol{\'o}rzano, The cell tracking challenge: 10 years of objective benchmarking, Nat. Methods 20~(7) (2023) 1010--1020.

\bibitem{Dey2023-ei}
N.~Dey, S.~Mazdak~Abulnaga, B.~Billot, E.~A. Turk, P.~Ellen~Grant, A.~V. Dalca, P.~Golland, {AnyStar}: Domain randomized universal star-convex {3D} instance segmentation (Jul. 2023).
\newblock \href {http://arxiv.org/abs/2307.07044} {\path{arXiv:2307.07044}}.

\bibitem{Scherr2020-ml}
T.~Scherr, K.~L{\"o}ffler, M.~B{\"o}hland, R.~Mikut, Cell segmentation and tracking using {CNN-based} distance predictions and a graph-based matching strategy, PLoS One 15~(12) (2020) e0243219.

\bibitem{Xie2021-tr}
E.~Xie, W.~Wang, Z.~Yu, A.~Anandkumar, J.~M. Alvarez, P.~Luo, {SegFormer}: Simple and efficient design for semantic segmentation with transformers, Adv. Neural Inf. Process. Syst. 34 (2021) 12077--12090.

\bibitem{Liu2022-ns}
Z.~Liu, H.~Mao, C.-Y. Wu, C.~Feichtenhofer, T.~Darrell, S.~Xie, A {ConvNet} for the 2020s (2022) 11976--11986\href {http://arxiv.org/abs/2201.03545} {\path{arXiv:2201.03545}}.

\bibitem{Xie2016-fp}
S.~Xie, R.~Girshick, P.~Doll{\'a}r, Z.~Tu, K.~He, Aggregated residual transformations for deep neural networks (2016) 1492--1500\href {http://arxiv.org/abs/1611.05431} {\path{arXiv:1611.05431}}.

\bibitem{Isensee2021-wd}
F.~Isensee, P.~F. Jaeger, S.~A.~A. Kohl, J.~Petersen, K.~H. Maier-Hein, {nnU-Net}: a self-configuring method for deep learning-based biomedical image segmentation, Nat. Methods 18~(2) (2021) 203--211.

\bibitem{Lee2022-by}
G.~Lee, S.~Kim, J.~Kim, S.-Y. Yun, {MEDIAR}: Harmony of {Data-Centric} and {Model-Centric} for {Multi-Modality} microscopy (Dec. 2022).
\newblock \href {http://arxiv.org/abs/2212.03465} {\path{arXiv:2212.03465}}.

\bibitem{Royer2023-ft}
L.~A. Royer, The future of bioimage analysis: a dialog between mind and machine, Nat. Methods 20~(7) (2023) 951--952.

\bibitem{Wu2023-au}
C.~Wu, S.~Yin, W.~Qi, X.~Wang, Z.~Tang, N.~Duan, Visual {ChatGPT}: Talking, drawing and editing with visual foundation models (Mar. 2023).
\newblock \href {http://arxiv.org/abs/2303.04671} {\path{arXiv:2303.04671}}.

\bibitem{Li2023-jw}
X.~Li, Y.~Zhang, J.~Wu, Q.~Dai, Challenges and opportunities in bioimage analysis, Nat. Methods 20~(7) (2023) 958--961.

\bibitem{Gao2021-ft}
Y.~Gao, M.~Zhou, D.~N. Metaxas, {UTNet}: A hybrid transformer architecture for medical image segmentation, in: Medical Image Computing and Computer Assisted Intervention -- {MICCAI} 2021, Springer International Publishing, 2021, pp. 61--71.

\bibitem{Wang2022-jp}
R.~Wang, T.~Lei, R.~Cui, B.~Zhang, H.~Meng, A.~K. Nandi, Medical image segmentation using deep learning: A survey, IET Image Proc. 16~(5) (2022) 1243--1267.

\bibitem{Kirillov2023-qn}
A.~Kirillov, E.~Mintun, N.~Ravi, H.~Mao, C.~Rolland, L.~Gustafson, T.~Xiao, S.~Whitehead, A.~C. Berg, W.-Y. Lo, P.~Doll{\'a}r, R.~Girshick, Segment anything (Apr. 2023).
\newblock \href {http://arxiv.org/abs/2304.02643} {\path{arXiv:2304.02643}}.

\bibitem{Caicedo2019-dc}
J.~C. Caicedo, A.~Goodman, K.~W. Karhohs, B.~A. Cimini, J.~Ackerman, M.~Haghighi, C.~Heng, T.~Becker, M.~Doan, C.~McQuin, M.~Rohban, S.~Singh, A.~E. Carpenter, Nucleus segmentation across imaging experiments: the 2018 data science bowl, Nat. Methods 16~(12) (2019) 1247--1253.

\bibitem{Archit2023-kl}
A.~Archit, S.~Nair, N.~Khalid, P.~Hilt, V.~Rajashekar, M.~Freitag, S.~Gupta, A.~Dengel, S.~Ahmed, C.~Pape, Segment anything for microscopy (Aug. 2023).

\bibitem{Paul-Gilloteaux2023-cw}
P.~Paul-Gilloteaux, Bioimage informatics: Investing in software usability is essential, PLoS Biol. 21~(7) (2023) e3002213.

\bibitem{Kemmer2023-ko}
I.~Kemmer, A.~Keppler, B.~Serrano-Solano, A.~Rybina, B.~{\"O}zdemir, J.~Bischof, A.~El~Ghadraoui, J.~E. Eriksson, A.~Mathur, Building a {FAIR} image data ecosystem for microscopy communities, Histochem. Cell Biol. 160~(3) (2023) 199--209.

\bibitem{Wilkinson2016-sq}
M.~D. Wilkinson, M.~Dumontier, I.~J.~J. Aalbersberg, G.~Appleton, M.~Axton, A.~Baak, N.~Blomberg, J.-W. Boiten, L.~B. da~Silva~Santos, P.~E. Bourne, J.~Bouwman, A.~J. Brookes, T.~Clark, M.~Crosas, I.~Dillo, O.~Dumon, S.~Edmunds, C.~T. Evelo, R.~Finkers, A.~Gonzalez-Beltran, A.~J.~G. Gray, P.~Groth, C.~Goble, J.~S. Grethe, J.~Heringa, P.~A.~C. 't~Hoen, R.~Hooft, T.~Kuhn, R.~Kok, J.~Kok, S.~J. Lusher, M.~E. Martone, A.~Mons, A.~L. Packer, B.~Persson, P.~Rocca-Serra, M.~Roos, R.~van Schaik, S.-A. Sansone, E.~Schultes, T.~Sengstag, T.~Slater, G.~Strawn, M.~A. Swertz, M.~Thompson, J.~van~der Lei, E.~van Mulligen, J.~Velterop, A.~Waagmeester, P.~Wittenburg, K.~Wolstencroft, J.~Zhao, B.~Mons, The {FAIR} guiding principles for scientific data management and stewardship, Sci Data 3 (2016) 160018.

\bibitem{Merkel2014-xs}
D.~Merkel, Docker: lightweight linux containers for consistent development and deployment, Linux J. 2014~(239) (2014) 2.

\bibitem{Zhou2023-dz}
Y.~Zhou, J.~Sonneck, S.~Banerjee, S.~D{\"o}rr, A.~Gr{\"u}neboom, K.~Lorenz, J.~Chen, {EfficientBioAI}: Making bioimaging {AI} models efficient in energy, latency and representation (Jun. 2023).
\newblock \href {http://arxiv.org/abs/2306.06152} {\path{arXiv:2306.06152}}.

\bibitem{Saraiva2023-nf}
B.~M. Saraiva, I.~M. Cunha, A.~D. Brito, G.~Follain, R.~Portela, R.~Haase, P.~M. Pereira, G.~Jacquemet, R.~Henriques, {NanoPyx}: super-fast bioimage analysis powered by adaptive machine learning (Aug. 2023).

\bibitem{Haase2020-pd}
R.~Haase, A.~Jain, S.~Rigaud, D.~Vorkel, P.~Rajasekhar, T.~Suckert, T.~J. Lambert, J.~Nunez-Iglesias, D.~P. Poole, P.~Tomancak, E.~W. Myers, Interactive design of {GPU-accelerated} image data flow graphs and cross-platform deployment using multi-lingual code generation (Nov. 2020).

\bibitem{Ouyang2023-sf}
W.~Ouyang, K.~W. Eliceiri, B.~A. Cimini, Moving beyond the desktop: prospects for practical bioimage analysis via the web, Front Bioinform 3 (2023) 1233748.

\bibitem{Von_Chamier2021-fp}
L.~von Chamier, R.~F. Laine, J.~Jukkala, C.~Spahn, D.~Krentzel, E.~Nehme, M.~Lerche, S.~Hern{\'a}ndez-P{\'e}rez, P.~K. Mattila, E.~Karinou, S.~Holden, A.~C. Solak, A.~Krull, T.-O. Buchholz, M.~L. Jones, L.~A. Royer, C.~Leterrier, Y.~Shechtman, F.~Jug, M.~Heilemann, G.~Jacquemet, R.~Henriques, Democratising deep learning for microscopy with {ZeroCostDL4Mic}, Nat. Commun. 12~(1) (2021) 2276.

\bibitem{Da_Veiga_Leprevost2017-cc}
F.~da~Veiga~Leprevost, B.~A. Gr{\"u}ning, S.~Alves~Aflitos, H.~L. R{\"o}st, J.~Uszkoreit, H.~Barsnes, M.~Vaudel, P.~Moreno, L.~Gatto, J.~Weber, M.~Bai, R.~C. Jimenez, T.~Sachsenberg, J.~Pfeuffer, R.~Vera~Alvarez, J.~Griss, A.~I. Nesvizhskii, Y.~Perez-Riverol, {BioContainers}: an open-source and community-driven framework for software standardization, Bioinformatics 33~(16) (2017) 2580--2582.

\bibitem{Bai2021-vb}
J.~Bai, C.~Bandla, J.~Guo, R.~Vera~Alvarez, M.~Bai, J.~A. Vizca{\'\i}no, P.~Moreno, B.~Gr{\"u}ning, O.~Sallou, Y.~Perez-Riverol, {BioContainers} registry: Searching bioinformatics and proteomics tools, packages, and containers, J. Proteome Res. 20~(4) (2021) 2056--2061.

\bibitem{Rubens2020-fr}
U.~Rubens, R.~Mormont, L.~Paavolainen, V.~B{\"a}cker, B.~Pavie, L.~A. Scholz, G.~Michiels, M.~Ma{\v s}ka, D.~{\"U}nay, G.~Ball, R.~Hoyoux, R.~Vandaele, O.~Golani, S.~G. Stanciu, N.~Sladoje, P.~Paul-Gilloteaux, R.~Mar{\'e}e, S.~Tosi, {BIAFLOWS}: A collaborative framework to reproducibly deploy and benchmark bioimage analysis workflows, Patterns (N Y) 1~(3) (2020) 100040.

\bibitem{Prigent2022-xg}
S.~Prigent, C.~A. Valades-Cruz, L.~Leconte, L.~Maury, J.~Salamero, C.~Kervrann, {BioImageIT}: Open-source framework for integration of image data management with analysis, Nat. Methods 19~(11) (2022) 1328--1330.

\bibitem{Ouyang2022-zw}
W.~Ouyang, F.~Beuttenmueller, E.~G{\'o}mez-de Mariscal, C.~Pape, T.~Burke, C.~Garcia-L{\'o}pez-de Haro, C.~Russell, L.~Moya-Sans, C.~de-la Torre-Guti{\'e}rrez, D.~Schmidt, D.~Kutra, M.~Novikov, M.~Weigert, U.~Schmidt, P.~Bankhead, G.~Jacquemet, D.~Sage, R.~Henriques, A.~Mu{\~n}oz-Barrutia, E.~Lundberg, F.~Jug, A.~Kreshuk, {BioImage} model zoo: A {Community-Driven} resource for accessible deep learning in {BioImage} analysis (Jun. 2022).

\bibitem{Ouyang2019-yw}
W.~Ouyang, F.~Mueller, M.~Hjelmare, E.~Lundberg, C.~Zimmer, {ImJoy}: an open-source computational platform for the deep learning era, Nat. Methods 16~(12) (2019) 1199--1200.

\bibitem{Bannon2021-oz}
D.~Bannon, E.~Moen, M.~Schwartz, E.~Borba, T.~Kudo, N.~Greenwald, V.~Vijayakumar, B.~Chang, E.~Pao, E.~Osterman, W.~Graf, D.~Van~Valen, {DeepCell} kiosk: scaling deep learning-enabled cellular image analysis with kubernetes, Nat. Methods 18~(1) (2021) 43--45.

\bibitem{Weisbart2023-mq}
E.~Weisbart, B.~A. Cimini, {Distributed-Something}: scripts to leverage {AWS} storage and computing for distributed workflows at scale, Nat. Methods 20~(8) (2023) 1120--1121.

\bibitem{Schmied2023-xy}
C.~Schmied, M.~S. Nelson, S.~Avilov, G.-J. Bakker, C.~Bertocchi, J.~Bischof, U.~Boehm, J.~Brocher, M.~T. Carvalho, C.~Chiritescu, J.~Christopher, B.~A. Cimini, E.~Conde-Sousa, M.~Ebner, R.~Ecker, K.~Eliceiri, J.~Fernandez-Rodriguez, N.~Gaudreault, L.~Gelman, D.~Grunwald, T.~Gu, N.~Halidi, M.~Hammer, M.~Hartley, M.~Held, F.~Jug, V.~Kapoor, A.~A. Koksoy, J.~Lacoste, S.~Le~D{\'e}v{\'e}dec, S.~Le~Guyader, P.~Liu, G.~G. Martins, A.~Mathur, K.~Miura, P.~Montero~Llopis, R.~Nitschke, A.~North, A.~C. Parslow, A.~Payne-Dwyer, L.~Plantard, R.~Ali, B.~Schroth-Diez, L.~Sch{\"u}tz, R.~T. Scott, A.~Seitz, O.~Selchow, V.~P. Sharma, M.~Spitaler, S.~Srinivasan, C.~Strambio-De-Castillia, D.~Taatjes, C.~Tischer, H.~K. Jambor, Community-developed checklists for publishing images and image analyses, Nat. Methods (Sep. 2023).

\bibitem{Hirling2023-eg}
D.~Hirling, E.~Tasnadi, J.~Caicedo, M.~V. Caroprese, R.~Sj{\"o}gren, M.~Aubreville, K.~Koos, P.~Horvath, Segmentation metric misinterpretations in bioimage analysis, Nat. Methods (Jul. 2023).

\bibitem{Maier-Hein2022-eb}
L.~Maier-Hein, A.~Reinke, P.~Godau, M.~D. Tizabi, F.~Buettner, E.~Christodoulou, B.~Glocker, F.~Isensee, J.~Kleesiek, M.~Kozubek, M.~Reyes, M.~A. Riegler, M.~Wiesenfarth, A.~Emre~Kavur, C.~H. Sudre, M.~Baumgartner, M.~Eisenmann, D.~Heckmann-N{\"o}tzel, A.~Tim~R{\"a}dsch, L.~Acion, M.~Antonelli, T.~Arbel, S.~Bakas, A.~Benis, M.~Blaschko, M.~Jorge~Cardoso, V.~Cheplygina, B.~A. Cimini, G.~S. Collins, K.~Farahani, L.~Ferrer, A.~Galdran, B.~van Ginneken, R.~Haase, D.~A. Hashimoto, M.~M. Hoffman, M.~Huisman, P.~Jannin, C.~E. Kahn, D.~Kainmueller, B.~Kainz, A.~Karargyris, A.~Karthikesalingam, H.~Kenngott, F.~Kofler, A.~Kopp-Schneider, A.~Kreshuk, T.~Kurc, B.~A. Landman, G.~Litjens, A.~Madani, K.~Maier-Hein, A.~L. Martel, P.~Mattson, E.~Meijering, B.~Menze, K.~G.~M. Moons, H.~M{\"u}ller, B.~Nichyporuk, F.~Nickel, J.~Petersen, N.~Rajpoot, N.~Rieke, J.~Saez-Rodriguez, C.~I. S{\'a}nchez, S.~Shetty, M.~van Smeden, R.~M. Summers, A.~A. Taha, A.~Tiulpin, S.~A. Tsaftaris, B.~Van~Calster, G.~Varoquaux, P.~F. J{\"a}ger,
  Metrics reloaded: Recommendations for image analysis validation (Jun. 2022).
\newblock \href {http://arxiv.org/abs/2206.01653} {\path{arXiv:2206.01653}}.

\end{thebibliography}

%% else use the following coding to input the bibitems directly in the
%% TeX file.

% \begin{thebibliography}{00}

% %% \bibitem{label}
% %% Text of bibliographic item

% \bibitem{}

% \end{thebibliography}
\end{document}